\documentclass{article}
\makeatletter
\@addtoreset{equation}{section}

\makeatother

\usepackage{graphicx}

\begin{document}

\begin{flushright}
Aug 2003

KEK-TH-909
\end{flushright}

\begin{center}

\vspace{5cm}

{\Large On Decay of Bulk Tachyons}

\vspace{2cm}

Takao Suyama \footnote{e-mail address : tsuyama@post.kek.jp}

\vspace{1cm}

{\it Theory Group, KEK}

{\it Tsukuba, Ibaraki 305-0801, Japan}

\vspace{4cm}

{\bf Abstract} 

\end{center}

We investigate a decay of a bulk tachyon with a Kaluza-Klein momentum in bosonic and Type 0 string 
theories compactified on $S^1$. 
Potential for the tachyon has a (local) minimum. 
A decay of the tachyon would lead the original theory to a strongly coupled theory. 
An endpoint of the decay would exist if the strong coupling limit exists and it is a stable theory. 

\newpage

\vspace{1cm}

\section{Introduction}

\vspace{5mm}

Recent investigations have provided an idea of how to understand the condensation of closed string 
tachyons which are localized in target space, and they have turned out to be similar to those of open 
string tachyons \cite{Sen}. 
An instability indicated by the localized tachyon is cured after the tachyon acquires a non-vanishing 
vev. 
Corresponding to a decay of branes in open string cases, the localized tachyon condensation deforms the 
background spacetime drastically \cite{APS}. 
Localized tachyons appear in a non-supersymmetric background, and generically it is conjectured to be 
deformed to a supersymmetric background by the tachyon condensation. 
The analyses have been done by using worldsheet RG flow 
\cite{Suyama}\cite{Vafa}\cite{HKMM}\cite{DGMS} and D-brane 
probes \cite{Yi}\cite{MT}. 
On the other hand, it is still difficult to understand the stabilization of an instability indicated by 
tachyons propagating in the bulk spacetime. 
There are conjectures on bulk tachyons in ten-dimensional tachyonic closed string theories 
\cite{CG}\cite{Suyama2}

In this paper, we would like to discuss a decay of a kind of bulk closed string tachyons. 
We will investigate a classical solution of a low energy effective theory consisting of the bulk 
tachyon and background fields, and deduce a result of the decay of the tachyon. 

To do this, we have to know about the shape of the potential for the bulk tachyon. 
For a massless field, its effective action including a potential term can be obtained from on-shell 
scattering amplitudes. 
However, since a zero momentum tachyon is an off-shell state, it is extremely difficult to know 
about the tachyon potential, see e.g. \cite{Banks}\cite{Zwiebach}. 

We will consider the following situation
\footnote{We noticed that a very similar calculation had been done in \cite{Klevanov}. 
We would like to thank I.Klevanov for informing us of the earlier work. }. 
Our interests are on bosonic string and Type 0 strings compactified on $S^1$. 
In such theories, bulk tachyons can have Kaluza-Klein momenta, and for appropriate value of the 
compactification radius, some of them can become massless. 
Therefore, it is possible to determine the effective potential for the massless `tachyon' in 25- or 
9-dimensional effective theory. 
The potential consists of non-derivative terms in lower dimensional sense, but it 
would contain contributions of derivatives with respect to the coordinate for the $S^1$ direction in 
higher dimensional sense. 
If the coefficients in the potential are finite, they would be still finite and have the same signs 
even when the radius is slightly changed so that the massless `tachyon' becomes really tachyonic. 
In this way, we can deduce the shape of the potential for the nearly-massless tachyons. 
We will show that the four-point coupling constant for the massless `tachyon' is positive and finite. 
Therefore, there is a (probably local) minimum of the potential for an appropriate value of the radius. 
It should be noted that we simply ignore tachyons with masses of order of the string scale, if exists. 

Then we will analyze a classical solution of the low energy effective theory, which describes a 
homogeneous decay of the tachyon. 
A typical behavior of the solution is that dilaton tend to become large without upper bound. 
This would indicate that the original theory becomes strongly coupled during the decay. 
There are some conjectures on the strong coupling behavior of bosonic string, which first appeared in 
\cite{Rey} and reconsidered later \cite{bosonicM}, and Type 0 
strings \cite{BG}, and both of them claim that the strong coupling limit is a stable theory. 
Thus, it would be natural to expect that the instability indicated by the bulk tachyon would be cured 
by that process described by the classical solution. 

This paper is organized as follows. 
In section \ref{boson}, we calculate the four-point coupling constant for tachyons in bosonic string 
theory. 
Then we analyze a classical solution of tachyon-graviton-dilaton system in section \ref{classical}. 
In section \ref{0A}, we extend our analysis to Type 0 string theories, and deduce an endpoint of the 
tachyon decay, according to the previous conjectures. 
Section \ref{dis} is devoted to discussion. 
Two appendices summarize spectra of some Type II and Type 0 orbifolds.

\vspace{1cm}

\section{Bulk tachyons in bosonic string theory}  \label{boson}

\vspace{5mm}

Let us start with the investigation of bulk closed string tachyons in bosonic string theory. 
We will calculate a four-point coupling constant for the tachyons in the low energy effective theory, 
which can be read off from a four-point scattering amplitude. 
We will follow conventions of \cite{Pol}. 

The four-point coupling constant can be obtained from the four-point scattering amplitude by the 
well-known procedure \cite{Bardacki}\cite{BH}, that is, by subtracting massless poles from the 
amplitude and 
taking the zero-momentum limit. 
For the case of tachyons, however, this procedure is not straightforward since one has to extend the 
amplitude to an off-shell region so as to take the limit. 
This has been done in the case of open string tachyons \cite{Bardacki}\cite{BH}. 
To avoid such difficulties, we consider the following setup. 
When the target space is compactified on $S^1$ with radius $R$, tachyons can have Kaluza-Klein 
momenta and their masses are 
\begin{equation}
M_n^2 = \left( \frac nR \right)^2-\frac4{\alpha'}.
\end{equation}
Thus, by tuning $R$, one of the Kaluza-Klein states, denoted as $|n\rangle$, can become massless, and 
then we can easily calculate the four-point coupling constant for the state $|n\rangle$. 
If the coupling constant is a non-zero and finite value, it would be still non-zero and finite even 
when the radius $R$ is slightly changed so that the state $|n\rangle$ becomes tachyonic. 
Therefore, in this way we can deduce from on-shell amplitudes the shape of the potential for the bulk 
closed string tachyon. 

\vspace{3mm}

The vertex operator of the Kaluza-Klein state is 
\begin{equation}
V_i(z,\bar{z}) = e^{iK_i\cdot X(z,\bar{z})}, \hspace{5mm} (i=1,2,3,4), 
\end{equation}
where $K_i=(k_{i\mu},\kappa_i),\ \mu=0,1,\cdots,24$. 
We choose that 
\begin{equation}
\kappa_1=-\kappa_2=\kappa_3=-\kappa_4=\frac nR,
\end{equation}
and fix the $SL(2,{\bf C})$ symmetry in the amplitude by setting $z_1=0,z_2=z,z_3=1,z_4\to\infty$. 
Then, the four-point scattering amplitude is, up to the delta function, 
\begin{eqnarray}
{\cal A}_4 
&=& \frac{8\pi ig_{st}^2}{\alpha'}\int_{\bf C} d^2z\ |z|^{-\alpha's/2-4}|1-z|^{-\alpha't/2-4}
     \nonumber \\
&=& \frac{16\pi^2 ig_{st}^2}{\alpha'}
    \frac{\Gamma\left(-\frac{\alpha'}4s-1\right)\Gamma\left(-\frac{\alpha'}4t-1\right)
          \Gamma\left(-\frac{\alpha'}4u+\frac{\alpha'}4M_{2n}^2\right)}
         {\Gamma\left(-\frac{\alpha'}4(s+t)-2\right)\Gamma\left(\frac{\alpha'}4s+2\right)
          \Gamma\left(\frac{\alpha'}4t+2\right)}.
     \label{boson4pt}
\end{eqnarray}
We have defined the Mandelstam variables of the compactified theory as follows, 
\begin{equation}
s=-(k_1+k_2)^2, \hspace{5mm} t=-(k_2+k_3)^2, \hspace{5mm} u=-(k_1+k_3)^2. 
\end{equation}

From now on, we set $R=|n|\sqrt{\alpha'}/2$ for which the state $|\pm n\rangle$ is massless, and thus 
we can take the limit $s,t,u\to0$ while keeping the on-shell condition. 
The amplitude (\ref{boson4pt}) has massless poles in the s- and the t-channel, but not in the 
u-channel since the mass of the lightest state propagating in this channel is $M_{2n}>0$. 
The pole structure of the product of gamma functions in (\ref{boson4pt}) is 
\begin{eqnarray}
& & \frac{\Gamma(-x-1)\Gamma(-y-1)\Gamma(x+y+3)}{\Gamma(-x-y-2)\Gamma(x+2)\Gamma(y+2)} \nonumber \\
&=& -\frac1x\pi x\cot(\pi x)\left( \frac{\Gamma(x+y+3)}{\Gamma(x+2)\Gamma(y+2)} \right)^2
    -\frac1y\pi y\cot(\pi y)\left( \frac{\Gamma(x+y+3)}{\Gamma(x+2)\Gamma(y+2)} \right)^2. \nonumber \\
&=& -\frac{(y+2)^2}x-\frac{(x+2)^2}y-8+O(x,y). 
\end{eqnarray}
The first two terms in the above expansion correspond to the exchange of massless closed string 
states. 
The tachyon-tachyon-massless vertex comes from the kinetic term of the tachyon 
\begin{equation}
-\int d^{26}x\ \sqrt{-G}e^{-2\Phi}G^{MN}\partial_MT^\dag\partial_NT
\end{equation}
via the dimensional reduction. 
We have denote a spacetime field corresponding to the state $|n\rangle$ as $T$. 
Note that $T$ is a complex scalar and the complex conjugate $T^\dag$ corresponds to the state 
$|-n\rangle$. 
The amplitude corresponding to the s-channel pole is, up to an overall factor, 
\begin{eqnarray}
\frac{((K_1-K_2)\cdot(K_3-K_4))^2}{(k_1+k_2)^2}
&=& -\frac{(2t+s+16/\alpha')^2}s \nonumber \\
&\propto& -\frac{(\alpha't/4+2)^2}{\alpha's/4}-2-\frac{\alpha'}4t-\frac{\alpha'}{16}s. 
\end{eqnarray}
By subtracting this and the t-channel poles from the amplitude (\ref{boson4pt}) and taking the 
zero-momentum limit, we obtain the following exact value of the four-point 
coupling constant $c_4$, 
\begin{eqnarray}
c_4
&=& i\lim_{s,t\to0}\left[ {\cal A}_4-\frac{16\pi^2ig_{st}^2}{\alpha'}
    \left(-\frac{(\alpha't/4+2)^2}{\alpha's/4}-\frac{(\alpha's/4+2)^2}{\alpha't/4}-4 \right) \right] 
   \nonumber \\
&=& \frac{64\pi^2g_{st}^2}{\alpha'}. 
\end{eqnarray}
Therefore we have obtained leading terms of the tachyon potential $V(T,T^\dag)$, 
\begin{equation}
V(T,T^\dag) = M_n^2T^\dag T+\frac{c_4}4(T^\dag T)^2
              +\cdots. 
\end{equation}
Here $\cdots$ indicates higher order terms in $T$ and $T^\dag$. 

Let us ignore the higher order terms for a while. 
Assume that $M_n^2<0$. 
Then the minimum of the potential is at $T=T_0$ where 
\begin{equation}
|T_0| = \sqrt{-\frac{2M_n^2}{c_4}}, \hspace{5mm} V(T_0,T_0^\dag)=-\frac{M_n^4}{c_4}. 
\end{equation}
Therefore, by taking $R$ to be close to the critical value $|n|\sqrt{\alpha'}/2$, the absolute value 
of $T_0$ can be taken to be arbitrarily small, so that a perturbative analysis of this minimum seems 
to be possible. 
(We will show in the next section that it is not the case.) 
Since $|T_0|$ can be arbitrarily small, contributions from higher order terms to the shape of the 
potential around $T=T_0$ would be negligible if $|M_n^2|$ is small enough. 

It may be possible to take a scaling limit which discards the higher order terms. 
The coefficient of the $k$-th order term in the potential $V(T,T^\dag)$ would be 
$c_k\sim g_{st}^{k-2}/\alpha'$. 
Therefore, by taking the following limit, 
\begin{equation}
\alpha'\to0, \hspace{5mm} g_{st}\to0, \hspace{5mm} c_4=\mbox{fixed}, 
\end{equation}
all coefficients $c_{k>4}$ vanish.  
However, what is important and also interesting in closed string tachyon condensation is its effect on 
background fields. 
Thus we take both $\alpha',g_{st}$ small but finite and $c_4=O(1)$, and just ignore all higher order 
terms in $T,T^\dag$ in the potential $V(T,T^\dag)$. 

We conclude that, for a suitable setup described above, the potential for a tachyon field 
$T$ has a minimum. 
It would be natural to think that the minimum describes an endpoint of condensation of the 
tachyon $T$. 
If so, the next question is to ask what kind of a theory the minimum describes. 
In the next section, we will investigate a time-dependent solution of the tachyon decay in 
terms of low energy field theory. 
We will find that, contrary to the naive expectation, the endpoint of the decay might not be a theory which is the expansion around the minimum. 
In addition, the endpoint might not be a static configuration, nor might be a perturbative one.

\vspace{1cm}

\section{Time-dependent tachyon decay}  \label{classical}

\vspace{5mm}

Since we have obtained leading two terms of the potential $V(T,T^\dag)$ for a nearly massless tachyon 
$T$, we can investigate its decay in terms of low energy field theory. 
We will set both $\alpha',g_{st}$ to be small but finite values, and neglect all higher order terms in 
$V(T,T^\dag)$. 
As mentioned before, these higher order terms are irrelevant for the presence of the minimum of the 
potential for small $M_n^2$, and thus results obtained below would be valid at least qualitatively. 
A naive expectation for the behavior of the classical solution for the decay would be that $T$ 
approaches to a non-zero value corresponding to the minimum of $V(T,T^\dag)$, but it will turn out 
that it is not the case. 

It should be noted that there are tachyons whose masses are of order of the string scale. 
We just omit them and concentrate on dynamics of the nearly massless tachyon $T$. 
Since the magnitude of the tachyon field $T$ would be small, it would be a good approximation to set 
the other tachyons to vanish. 
We will discuss in the next section that there is a situation without such a subtlety. 

\vspace{3mm}

The action of the low energy effective theory in 25 dimensions is 
\begin{eqnarray}
S &=& S_G+S_T, \\
S_G &=& \frac1{2\kappa^2}\int d^{25}x\sqrt{-G}e^{-2\Phi}\left[ \ 
         R+4G^{\mu\nu}\partial_\mu\Phi\partial_\nu\Phi \ \right], \\
S_T &=& -\int d^{25}x\sqrt{-G}e^{-2\Phi}\left[ \ 
         G^{\mu\nu}\partial_\mu T^\dag\partial_\nu T+V(T,T^\dag) \ \right], 
\end{eqnarray}
where 
\begin{equation}
V(T,T^\dag) = M_n^2T^\dag T+\frac{c_4}4(T^\dag T)^2. 
\end{equation}
There should be some comments on massless fields which are omitted. 
The NS-NS B-field and the graviphoton field are omitted just for simplicity. 
In fact, the graviphoton field will get a mass if the Kaluza-Klein state $T$ acquires a vev, 
and thus the vanishing configuration would be a natural classical solution. 
The omission of $G_{25,25}$ is due to a technical reason. 
The vev of this field parametrizes the radius of the $S^1$, and thus effects of the tachyon 
decay on the radius would be caused by an interaction between $T$ and $G_{25,25}$. 
The problem is that we cannot determine the $R$-dependence of $c_4$, so that the interaction between 
$T$ and $G_{25,25}$ is unclear. 
Therefore, we fix the radius and ignore the dynamics of $G_{25,25}$. 
It would be interesting to determine $c_4$ for general $R$ in another method. 

Below, we will work in the Einstein frame. 
The action is, 
\begin{eqnarray}
S &=& \frac1{\kappa^2}\int d^Dx\sqrt{-G}
      \left[ \frac12R-\frac12G^{\mu\nu}\partial_\mu\phi\partial_\nu\phi 
     -G^{\mu\nu}\partial_\mu \tau^\dag\partial_\nu \tau
           -e^{\beta\phi}\kappa^2V(\tau/\kappa,\tau^\dag/\kappa) \right],
\end{eqnarray}
where we have rescaled as $\phi=\beta\Phi, \beta=\sqrt{4/(D-2)}, \tau=\kappa T$. 
The field equations are 
\begin{eqnarray}
R_{\mu\nu}-\frac12G_{\mu\nu}R &=& T_{\mu\nu}, \\
-G^{\mu\nu}\nabla_\mu\nabla_\nu\phi+\beta e^{\beta\phi}\kappa^2V(\tau/\kappa,\tau^\dag/\kappa) 
&=& 0, \\
-G^{\mu\nu}\nabla_\mu\nabla_\nu \tau+e^{\beta\Phi}\kappa^2\frac{\partial V}{\partial \tau^\dag} 
&=& 0,
\end{eqnarray}
where 
\begin{eqnarray}
T_{\mu\nu} 
&=& \partial_\mu\phi\partial_\nu\phi
   -\frac12G_{\mu\nu}G^{\rho\sigma}\partial_\rho\phi\partial_\sigma\phi    \nonumber \\
& &+\partial_\mu \tau^\dag\partial_\nu \tau+\partial_\nu \tau^\dag\partial_\mu \tau
   -G_{\mu\nu}G^{\rho\sigma}\partial_\rho \tau^\dag\partial_\sigma \tau
   -G_{\mu\nu}e^{\beta\phi}\kappa^2V(\tau/\kappa,\tau^\dag/\kappa). 
\end{eqnarray}

We would like to investigate a spatially homogeneous tachyon decay, and thus we assume that 
\begin{equation}
ds^2 = -dt^2+a(t)^2\delta_{kl}dx^kdx^l, \hspace{5mm} \phi=\phi(t), \hspace{5mm} \tau=\tau(t). 
\end{equation}
We also assume for simplicity that $\tau$ is restricted to be real. 
Note that this is just a gauge fixing condition with residual ${\bf Z}_2$ gauge symmetry. 
Then, after a suitable rescaling of the time variable $t$, the field equations are reduced to 
\begin{eqnarray}
H^2
&=& \frac{2}{(D-1)(D-2)}\left( \frac12\dot{\phi}^2+\dot{\tau}^2
    +e^{\beta\phi}{\cal V}(\tau) \right), 
   \label{H2} \\
\ddot{\phi} &=& -(D-1)H\dot{\phi}-\beta e^{\beta\phi}{\cal V}(\tau), 
   \label{EOMs} \\
\ddot{\tau} &=& -(D-1)H\dot{\tau}-\frac12e^{\beta\phi}\frac{d{\cal V}}{d\tau}, 
   \label{ddottau}
\end{eqnarray}
where $H=\dot{a}/a$ and 
\begin{equation}
V(\tau) = \mu^2\tau^2+\frac14\tau^4, \hspace{5mm} \mu^2 = \frac{\kappa^2}{c_4}M_n^2. 
\end{equation}

One may worry that the first equation (\ref{H2}) may lead to an unphysical solution since 
${\cal V}(\tau)$ 
can be negative. 
However, one can show that 
\begin{eqnarray}
\frac{d^2H^2}{dt^2} &=& 2\left( \frac{1}{D-2} \right)^2(\dot{\phi}^2+2\dot{\tau}^2)^2
                       -\frac{2}{D-2}H\frac{d}{dt}(\dot{\phi}^2+2\dot{\tau}^2), 
\end{eqnarray}
since 
\begin{equation}
\frac{dH}{dt} = -\frac{1}{D-2}(\dot{\phi}^2+2\dot{\tau}^2), 
    \label{Hubble} \\
\end{equation}
is derived from the Einstein equation. 
Thus the second derivative of $H^2$ is generically positive when $H=0$. 
This means that, as long as 
one starts with a meaningfull initial conditions, $H^2$ does not become negative, as it should be. 

\begin{figure}[htbp]
\rotatebox{-90}{
\includegraphics[width=.5\linewidth]{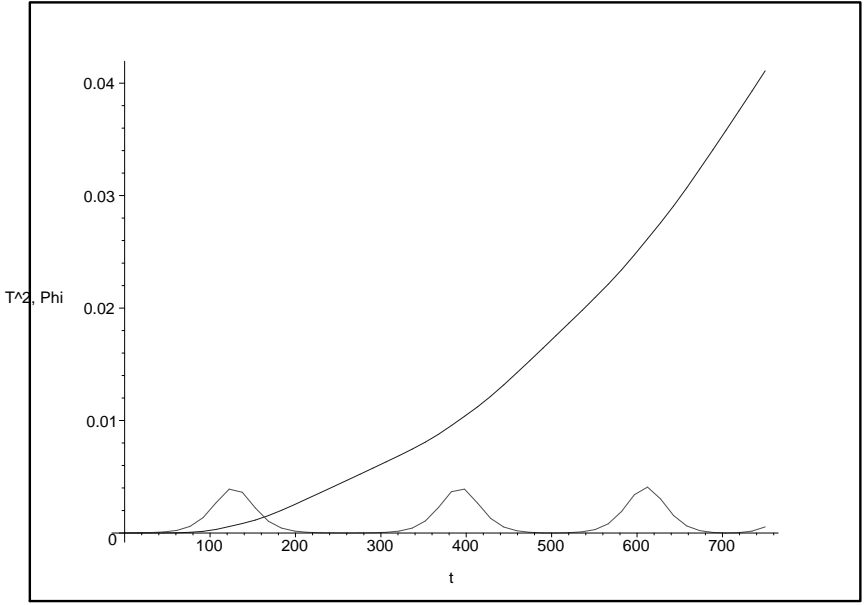}
}
\caption{A typical solution} \label{fig}
\end{figure}

A solution of the equations of motion (\ref{H2})-(\ref{ddottau}) can be obtained numerically. 
A set of natural initial consitions are 
\begin{eqnarray}
&& \phi(0)=0, \hspace{5mm} \dot{\phi}(0)=0, \nonumber \\
&& \tau(0)=\tau_i, \hspace{5mm} \dot{\tau}(0)=v_i, \hspace{5mm} 
\mbox{such that} \hspace{5mm} H=0. 
 \nonumber 
\end{eqnarray}
One may take $\phi(0)=\phi_0\ne0$, but $\phi_0$ can be absorbed by a suitable rescaling of $t$. 
One can see, from (\ref{Hubble}), that for these initial conditions $H$ is always negative for $t>0$. 
Therefore the gravity does not work as friction, and as a result, oscillation of $\tau$ will not be 
dumped.  
An example of the numerical solution is shown in figure \ref{fig}. 

A typical property of the solution is that the dilaton $\phi$ grows, probabry, without upper bound. 
This means that the tachyon decay would lead the original theory to a strongly coupled theory. 
$|H|$ and the scalar curvature also grow with time, but their evolutions are slower than that of the 
dilaton, at least in the early stage of the decay. 
In fact, $|H|$ and the scalar curvature is nearly zero in a range of time shown in figure \ref{fig}. 

One can understand this behavior by a simple order estimate. 
Suppose that $\dot{\tau}\sim \varepsilon$. 
Typically $\varepsilon\sim \mu^2$. 
Then as long as $\dot{\phi}$ is of order of $\varepsilon$ or smaller, one obtains 
$\dot{H}\sim\varepsilon^2$ and $H^2\sim\varepsilon^4$. 
Therefore $R\sim\varepsilon^2$ and thus this is negligible. 
Note that the typical scale of this system would be $\sqrt{\varepsilon}$. 
Since $\ddot{\phi}\sim\varepsilon^2$, $\phi$ can grow while $\dot{\phi}$ is kept small. 

It would be natural to expect that the tachyon decay may have an endpoint if the strong 
coupling limit of the theory exists and if it is a stable theory. 
If the scalar curvature becomes large after the dilaton becomes large enough, one would have to take 
into account corrections due to large curvature of the spacetime in order to deduce the endpoint of 
the decay. 
The strong coupling limit of bosonic string theory has been discussed in \cite{Rey}\cite{bosonicM}. 
In the next section, we repeat the same analysis for bulk tachyons in Type 0 theory compactified on 
$S^1$ and its orbifold. 
Since there is a conjecture on the strong coupling limit of Type 0 theory \cite{BG}, 
according to this conjecture, one could deduce an endpoint of the tachyon decay in Type 0 
theory. 

Note that we have plotted $\tau^2$ rather than $\tau$ since $\tau$ is originally a charged scalar $T$, 
and a gauge invariant operator is $T^\dag T$, not $T$. 
Clearly $\tau^2$ has a non-zero time average which would correspond to a non-vanishing vev of the 
tachyon.

\vspace{1cm}

\section{Bulk tachyons in Type 0 theory}  \label{0A}

\vspace{5mm}

We will show that the analysis done in previous sections can be extended to the case of Type 0 
theory. 
See also \cite{Klevanov}. 

First we consider Type 0 theory compactified on $S^1$ with radius $R$. 
The mass formula for Kaluza-Klein states without oscillator excitations is 
\begin{equation}
M_n^2 = \left( \frac nR \right)^2-\frac 2{\alpha'}. 
\end{equation}
When $R=|n|\sqrt{\alpha'/2}$, the states $|\pm n\rangle$ with $\pm n$ units of Kaluza-Klein momentum 
become massless. 
Vertex operators of the states are 
\begin{eqnarray}
V_{1,3} &=& e^{-\phi-\bar{\phi}}e^{iK_{1,3}\cdot X}, \\
V_{2,4} &=& (K_{2,4}\cdot\psi)(K_{2,4}\cdot\tilde{\psi})e^{iK_{2,4}\cdot X}, 
\end{eqnarray}
for (-1,-1) and (0,0) pictures, respectively. 
We choose their Kaluza-Klein momenta and the way of fixing $SL(2,{\bf C})$ symmetry as before. 
Then the resulting four-point amplitude is, up to an overall positive constant and the delta function, 
\begin{eqnarray}
{\cal A}_4 
&=& (K_2\cdot K_4)^2\int_{\bf C}d^2z\ |z|^{\alpha'K_1\cdot K_2}|1-z|^{\alpha'K_2\cdot K_3}
      \nonumber \\
&=& 8\pi\left( \frac{\alpha'}4(s+t)+1 \right)^2
    \frac{\Gamma\left(-\frac{\alpha'}4s\right)\Gamma\left(-\frac{\alpha'}4t\right)
          \Gamma\left(\frac{\alpha'}4(s+t)+1\right)}
         {\Gamma\left(-\frac{\alpha'}4(s+t)\right)\Gamma\left(\frac{\alpha'}4s+1\right)
          \Gamma\left(\frac{\alpha'}4t+1\right)}. 
   \label{Type0amp}
\end{eqnarray}

One can check the overall sign by examining the pole structure of the amplitude, since (\ref{Type0amp}) 
should have a similar structure to that for bosonic string theory. 
This is because R-R fields do not propagate in any channel. 
The relevant part of the amplitude can be rewritten as before, 
\begin{eqnarray}
& & (x+y+1)^2\frac{\Gamma(-x)\Gamma(-y)\Gamma(x+y+1)}{\Gamma(-x-y)\Gamma(x+1)\Gamma(t+1)} \nonumber \\
&=& -\frac1x\pi x\cot(\pi x)(x+y+1)^2\left( \frac{\Gamma(x+y+1)}{\Gamma(x+1)\Gamma(y+1)} \right)^2 
    \nonumber \\
& & -\frac1y\pi y\cot(\pi y)(x+y+1)^2\left( \frac{\Gamma(x+y+1)}{\Gamma(x+1)\Gamma(y+1)} \right)^2 
    \nonumber \\
&=& -\frac{(y+1)^2}x-\frac{(x+1)^2}y-4+O(x,y). 
\end{eqnarray}
One can see that the pole structure is almost the same as before (the difference comes from the 
difference of $R$), and thus this confirms the corresctness of 
the overall sign of (\ref{Type0amp}). 
By subtracting an appropriate quantity corresponding to the s- and t-channel poles, 
\begin{equation}
-\frac{(y+1)^2}x-\frac{(x+1)^2}y-2,
\end{equation}
one can show that the four-point coupling constant $c_4$ is agian positive and finite. 
This indicates that potential for a field $T$ corresponding to the state $|n\rangle$ would have a 
minimum when $R$ is slightly changed so that $T$ becomes slightly tachyonic. 

\vspace{3mm}

The analysis of time-dependent classical solution of the low energy effective theory is almost the 
same as in the bosonic case, and thus we conclude that the theory into which the Type 0 theory decays 
would be a strongly coupled theory. 
There is a conjecture on the strong coupling limit of Type 0 theory \cite{BG}. 
This conjecture claims that the strong coupling limit of Type 0A theory in ten dimensions is equivalent 
to an eleven-dimensional M-theory. 
More precisely, Type 0A theory with a finite string coupling $g_{st}$ is conjectured to be equivalent 
to M-theory compactified on 
$S^1$ with a twisted boundary condition for fermions in the $S^1$ direction. 
The radius of the $S^1$ is related to $g_{st}$, and the strong coupling limit corresponds to 
decompactification limit in M-theory. 
We summarize arguments on this duality in appendix \ref{dual}. 
According to this conjecture, the theory after the tachyon decay would be a Type IIA theory since we 
have started with Type 0 theory compactified on $S^1$. 
However, this would not be the supersymmetric Type IIA vacuum because of the presence of the tachyon 
vev. 
A state or a field which has the vev in Type IIA picture would be related to D0-branes since the 
tachyon is a Kaluza-Klein state in M-theory picture, but properties of them are unclear at present. 

\vspace{3mm}

A more interesting case is an orbifold of Type 0 theory compactified on $S^1$. 
Let us consider a Type 0 theory compacified on $S^1$ with orbifolding by $t^{1/2}\cdot(-1)^{F_R}$. 
Here $t^{1/2}$ is an operator which produces the shift along the $S^1$ by $\pi R$, and $F_R$ counts 
the number of right-moving worldsheet fermions. 
We will denote this theory as Type 0 theory on $S^1/t^{1/2}\cdot(-1)^{F_R}$. 
It is known that Type 0A theory on $S^1/t^{1/2}\cdot(-1)^{F_R}$ is T-dual to Type IIB theory 
on $S^1/t^{1/2}\cdot(-1)^{F_s}$, where $F_s$ is the spacetime fermion number operator. 
In addition, the latter is equivalent to Type IIB theory compactified on Melvin background with 
$2\pi$ rotation \cite{CG}. 
In this Type 0 orbifold, a bulk tachyon without Kaluza-Klein momentum is projected out since 
$(-1)^{F_R}|0\rangle=-|0\rangle$. 
Therefore, if we take $R$ to be slightly above $\sqrt{\alpha'/2}$, in the theory there are massless 
fields and a nearly-massless bulk tachyon, while all other fields have masses of order of the string 
scale, 
and thus there is no subtlety mentioned before in analyzing low energy effective theory. 

The tachyon decay would lead the Type 0 orbifold to M-theory on $S^1/t^{1/2}\cdot(-1)^{F_R}$. 
To make the situation clear, let us compactify one more direction on $S^1$ and regard this $S^1$ as 
the eleventh direction. 
Then the resulting theory is Type IIA theory on $S^1/t^{1/2}\cdot(-1)^{F_R}$. 
Mass spectrum of this theory is summarized in appendix \ref{spectra}. 
This theory is not supersymmetric, but perturbatively stable. 
Therefore, it is tempting to conjecture that an endpoint of the decay would be described by this 
Type II orbifold. 
Tachyon condensation in Melvin background with a fractional rotation has been investigated 
\cite{Suyama}\cite{DGMS} and 
it is claimed that the endpoint is generically the flat supersymmetric Type II vacuum. 
Then a naive guess is that it would be also the case for Melvin background with $2\pi$ rotation. 
If this guess is correct, the Type II orbifold discussed above might be a metastable vacuum. 
This metastable vacuum might not be able to be reached for general $R$ since in such cases higher 
order terms in tachyon 
fields would not be negligible and the minimum of the potential discussed so far would disappear. 
It is very interesting to clarify this issue.

\vspace{1cm}

\section{Discussion}  \label{dis}

\vspace{5mm}

We have investigated bulk tachyons in various closed string theory. 
We concentrated on tachyons with a Kaluza-Klein momentum, which can become massless at a 
special radius of a compactified direction. 
Such a situation enables us to extract information of the four-point coupling constant of the tachyons 
from an on-shell scattering amplitude. 
The resulting tachyon potential has a (local) minimum. 
Then we analyzed a classical solution for field equations of the tachyon, graviton and dilaton system. 
A typical property of the solution is that the dilaton grows with time, while the tachyon oscillates 
within a finite range and the curvature is almost zero, at least at the initial stage of the tachyon 
decay. 
This result led us to a conjecture on the tachyon decay; the endpoint of the decay would be a 
theory which describes the strong coupling limit of the original theory. 
Since conjectured strong coupling limits of bosonic string and Type 0 string are stable, it would be 
natural to expect that endpoints of the decay for the theories would exist. 

\vspace{3mm}

There may be subtleties for the late time behavior of the low energy fields in the tachyon decay. 

The first one is on the magnitude of the curvature. 
Since the curvature may become large when $\dot{\phi},\dot{\tau}$ become large, the theory describing 
an endpoint of the decay may have a curved background geometry. 
It is not clear whether the above statement makes sense, since the low energy field theory 
approximation employed in section \ref{classical} would break down after the dilaton becomes large. 
At least, it would be possible to expect that, by going into a strong coupling rigon, the intrinsic 
instability (the presence of tachyons) would be cured by the tachyon dynamics, and the problem of 
finding an endpoint of the decay would be reduce to an analysis of a non-trivial background in a stable 
theory. 

The second is on the amplitude of the tachyon oscillation. 
We have only considered the tachyon potential in the vicinity of $T=0$. 
If the amplitude of the oscillation becomes large so that $T$ escapes from the minimum, the above 
discussion would be meaningless, and the endpoint of the decay, if exists, would be completely 
different. 
However, we expect that $T$ would be confined within a finite range even in a later time, since the 
potential $e^{\beta\phi}{\cal V}(\tau)$ becomes steeper and thus the potential barrier becomes larger 
when the dilaton becomes larger. 

\vspace{3mm}

Our conjectures on the endpoints of the tachyon decays strongly depend on the validity of the 
conjectures on strong 
coupling limits of non-supersymmtric string theories. 
However, due to the lack of the powerful tools of supersymmetry, the latter conjectures are not yet 
convincing enough. 
Therefore, it is desired to refine the arguments on the conjectures to gain insights into the strong 
coupling behavior of the non-supersymmetric theories. 

\vspace{3mm}

It seems interesting to compare our claims on bulk tachyons with those on open string tachyons. 
We conjectured that a decay of a bulk tachyons with a non-zero Kaluza-Klein momentum could be deduced 
from a low energy effective theory analysis, but a decay of the zero mode tachyon is still out of 
reach of our understanding. 
Similarly in open string tachyon condensation, a tachyon with a Kaluza-Klein momentum leads a 
D25-brane to a lower-dimensional brane, and this phenomenon is well-understood. 
However, the condensation of the zero mode tachyon, which would describe the disappearance of the 
D25-brane, is more difficult to analyze, compared with the above case. 
This analogy might suggest that it would be necessary to employ techniques other than those in this 
paper for the investigation of decay of the zero mode bulk tachyon. 

\vspace{3mm}

It would be interesting to extend our analysis to various non-supersymmetric heterotic strings 
\cite{Suyama2}. 
Since in such theories tachyons are in general charged under gauge fields, and thus decays of the 
tachyons would be more complicated. 
It would be also possible to do the same analysis for theories with localized closed string tachyons, 
but in such theories, scattering amplitudes will be complicated functions and thus explicit 
calculations would be difficult. 

\vspace{1cm}

{\bf {\Large Acknowledgements}}

\vspace{5mm}

We would like to thank T.Asakawa, A.Dhar, Y.Kitazawa, S.Matsuura, H.Shin, P.Yi, E.Witten for valuable 
discussions.

\appendix

\newpage

\vspace{1cm}

\section{Strong coupling limit of Type 0A theory}  \label{dual}

\vspace{5mm}

In this appendix, we summarize arguments on a conjecture on the strong coupling limit of Type 0A 
theory in ten dimensions \cite{BG}. 

Consider Type IIA on $S^1/t^{1/2}\cdot(-1)^{F_s}$ whose definitions and notations are explained 
in section \ref{0A}. 
Denote the radius of the $S^1$ as $R$. 
Mass spectrum of this theory is as follows, 
\begin{equation}
\begin{array}{rll}
\mbox{untwisted sector} : & (NS+,NS+;2{\bf Z},{\bf Z}) & (\ R+\ ,\ R-\ ;2{\bf Z},{\bf Z}) \\
                   & (NS+,\ R-\ ;2{\bf Z}+1,{\bf Z}) & (\ R+\ ,NS+;2{\bf Z}+1,{\bf Z}) \\
                   &                   &                   \\
\mbox{twisted sector} :   & (NS-,NS-;2{\bf Z},{\bf Z}+1/2) & (\ R-\ ,\ R+\ ;2{\bf Z},{\bf Z}+1/2) \\
                   & (NS-,\ R+\ ;2{\bf Z}+1,{\bf Z}+1/2) & (\ R-\ ,NS-;2{\bf Z}+1,{\bf Z}+1/2) 
\end{array}
\end{equation}
where, for example, $(NS+,NS+;2{\bf Z},{\bf Z})$ represents a set of states in the NS-NS sector with 
$(-1)^{F_L}=(-1)^{F_R}=+1$, even Kaluza-Klein momentum numbers and integer winding numbers. 
One can see that the spectrum of bosonic states is almost equivalent to that of Type 0A theory on 
$S^1$ when $R$ is small so that the winding states are almost degenerate. 
In this case, all other states are fermionic with masses of order $1/R$. 

In \cite{BG}, it is argued that D0-branes in Type 0A theory could form bound states which would be 
spacetime fermions. 
Then Type IIA on $S^1/t^{1/2}\cdot(-1)^{F_s}$ may be regarded as Type 0A theory on $S^1$ including 
D-branes, by relating the fermionic states in the former to the D0 bound states in the latter. 
The conjectured equivalence is the following, 

\vspace{2mm}

\begin{center}

M on $S^1\times (S^1/t^{1/2}\cdot(-1)^{F_s})$ = Type 0A on $S^1$. 

\end{center}

\vspace{2mm}

Denote the radius of the trivial $S^1$ in the LHS as $R_1$, the other radius as $R_2$, the radius in 
the RHS as $R_0$ and the string coupling of Type 0A as $g_0$. 
The conjectured relations between them are 
\begin{equation}
g_0 \sim R_2^{3/2}, \hspace{5mm} R_0 \sim \sqrt{R_2}R_1. 
\end{equation}

The limit $R_1,R_2\to0$ may result in the equivalence between Type IIA on 
$S^1/t^{1/2}\cdot(-1)^{F_s}$ and Type 0A on $S^1$. 
Note that it would be difficult to show quantitative evidence for the relation between the fermionic 
states in the Type IIA orbifold and the D0 bound states in the Type 0A theory, since, for example, 
masses of them would receive a large quantum correction due to the absence of supersymmetry. 

Another limit $R_0\to\infty$ while $g_0$ is fixed corresponds to $R_1\to\infty$ and $R_2$ is fixed. 
This means the following equivalence 

\vspace{2mm}

\begin{center}

M on $S_1/t^{1/2}\cdot(-1)^{F_s}$ = Type 0A in ten dimensions. 

\end{center}

\vspace{2mm}

Note that the LHS has a geometric realization, that is, it is equivalent to M-theory on Melvin 
background with $2\pi$ rotation \cite{CG}.

\vspace{1cm}

\section{Spectra of Type II and Type 0 orbifolds}  \label{spectra}

\vspace{5mm}

We have shown in the previous appendix the spectrum of Type IIA on $S^1/t^{1/2}\cdot(-1)^{F_s}$. 
The spectrum of Type IIB on $S^1/t^{1/2}\cdot(-1)^{F_s}$ can then be easily obtained by changing the 
eigenvalue of $(-1)^{F_R}$ in $(NS,R),(R,R)$ sectors. 

Then we consider Type 0B on $S^1/t^{1/2}\cdot(-1)^{F_R}$. 
\begin{equation}
\begin{array}{rll}
\mbox{untwisted sector :} & (NS+,NS+;2{\bf Z},{\bf Z}) & (\ R+\ ,\ R+\ ;2{\bf Z},{\bf Z}) \\
                          & (NS-,NS-;2{\bf Z}+1,{\bf Z}) & (\ R-\ ,\ R-\ ;2{\bf Z}+1,{\bf Z}) \\
                          &                    &                          \\
\mbox{twisted sector :}   & (NS+,\ R+\ ;2{\bf Z},{\bf Z}+1/2) & (\ R+\ ,NS+;2{\bf Z}.{\bf Z}+1/2) \\
                          & (NS-,\ R-\ ;2{\bf Z}+1,{\bf Z}+1/2) & (\ R-\ ,NS-;2{\bf Z}+1,{\bf Z}+1/2) 
\end{array}
\end{equation}
The momentum and the winding numbers can be redefined by changing the radius as follows, 
\begin{equation}
(m,w)_R = \left(\frac m2,2w\right)_{R/2}. 
\end{equation}
Then, one can see that the spectrum after the above redefinition is T-dual to that of Type IIA on 
$S^1/t^{1/2}\cdot(-1)^{F_s}$, 
since T-duality makes changes in the egenvalue of $(-1)^{F_R}$ in $(NS,R),(R,R)$ sectors and exchanges 
the momentum and the winding numbers. 

\vspace{3mm}

Next we consider Type IIA on $S^1/t^{1/2}\cdot(-1)^{F_R}$. 
The spectrum is 
\begin{equation}
\begin{array}{rll}
\mbox{untwisted sector :} & (NS+,NS+;2{\bf Z},{\bf Z}) & (\ R+\ ,NS+;2{\bf Z},{\bf Z}) \\
                          & (NS+,\ R-\ ;2{\bf Z}+1,{\bf Z}) & (\ R+\ ,\ R-\ ;2{\bf Z}+1,{\bf Z}) \\
                          &                    &                          \\
\mbox{twisted sector :}   & (NS+,\ R+\ ;2{\bf Z},{\bf Z}+1/2) & (\ R+\ ,\ R+\ ;2{\bf Z}.{\bf Z}+1/2) \\
                          & (NS+,NS-;2{\bf Z}+1,{\bf Z}+1/2) & (\ R+\ ,NS-;2{\bf Z}+1,{\bf Z}+1/2) 
\end{array}
\end{equation}
One can see that this spectrum does not contain tachyons at all since there is no $(NS-,NS-)$ sector. 
Therefore, this theory is stable at least perturbatively.

\end{document}